\documentclass[A4paper,notitlepage,longbibliography,aps,twocolumn,prl]{revtex4-1}
\usepackage{graphicx}
\graphicspath{{figures/}}
\usepackage{amsfonts}
\usepackage{amssymb}
\usepackage{mathrsfs}
\usepackage{soul}
\usepackage[usenames, dvipsnames]{color}
\usepackage[colorlinks=true,citecolor=blue,urlcolor  = purple]{hyperref}

\usepackage{tikz}

\usepackage[mathcal]{euscript}

\usepackage{amsmath}

\usepackage{pdfpages}

\renewcommand{\d}{\mathrm{d}}
\newcommand{\Tr}{\operatorname{Tr}}

\newcommand{\ket}[1]{\left| #1 \right\rangle}
\newcommand{\bra}[1]{\left\langle #1 \right|}

\newcommand{\ketbra}[2]{| #1 \rangle \langle #2 |}

\newcommand{\kommentar}[1]{}

\usepackage{amsthm}

\newtheorem*{lemma*}{Lemma}
\newtheorem*{corollary*}{Corollary}
\theoremstyle{remark}

\usepackage{graphicx}

\usepackage{float}
\usepackage{color}
\definecolor{npurple}{rgb}{0.3,0,0.6}

\newcommand{\M}{\mathcal{M}}

\newcommand{\II}{\openone}
\newcommand{\I}{\openone}

\newcommand{\CC}{\mathbb{C}}

\newcommand{\K}{\mathcal{K}}

\newcommand{\lspan}{\operatorname{span}}

\newcommand{\PVM}{\mathrm{PVM}}
\newcommand{\POVM}{\mathrm{POVM}}

\newcommand{\red}[1]{{#1}}

\definecolor{mygray}{gray}{0.6}

\makeatletter
\AtBeginDocument{\let\LS@rot\@undefined}
\makeatother

\begin{document}


\title{Some quantum measurements with three outcomes can reveal nonclassicality \\ where all two-outcome measurements fail to do so}
\date{\today}

\author{H. Chau Nguyen}
\email{chau.nguyen@uni-siegen.de}
\affiliation{Naturwissenschaftlich-Technische Fakult\"at, Universit\"at Siegen,
Walter-Flex-Stra{\ss}e 3, 57068 Siegen, Germany}

\author{Otfried G\"{u}hne}
\email{otfried.guehne@uni-siegen.de}
\affiliation{Naturwissenschaftlich-Technische Fakult\"at, Universit\"at Siegen,
Walter-Flex-Stra{\ss}e 3, 57068 Siegen, Germany}

\begin{abstract}
Measurements serve as the intermediate communication layer between 
the quantum world and our classical perception. So, the question 
which measurements efficiently extract information from quantum 
systems is of central interest. Using quantum steering as 
a nonclassical phenomenon, we show that there are instances, where
the results of all two-outcome measurements can be explained in a 
classical manner, while the results of some three-outcome measurements 
cannot. This points at the important role of the number of outcomes 
in revealing the nonclassicality hidden in a quantum system.  
Moreover, our methods allow to improve the understanding of quantum 
correlations by delivering novel criteria for quantum steering and 
improved ways to construct local hidden variable models. 
\end{abstract}

\maketitle


{\it Introduction.---}
It is widely believed that, at the fundamental level, our world behaves 
according to the laws of quantum mechanics, although we can only perceive 
it classically~\cite{Zurek2003a}. In fact, realizing the hidden potential 
of quantum mechanical systems in information processing has ignited the burst 
of quantum information and quantum computation during the last years~\cite{Nielsen2010a}. 
To transfer the quantum mechanical concepts to that of our familiar classicality, 
quantum measurements are required~\cite{Schlosshauer2007a}. The question how 
to use quantum measurements to interact efficiently with quantum mechanical 
systems is thus of central interest in quantum information 
theory~\cite{Schlosshauer2007a}. 

In 1964, Bell found that measurements performed locally on a bipartite quantum 
system can yield results which cannot be explained with a classical intuition
based on the assumptions of locality and realism~\cite{Einstein1935a, Bell1964a}. 
This phenomenon manifests itself as the violation of Bell inequalities, and a 
famous example of such an inequality is the Clauser-Horne-Shimony-Holt (CHSH) 
inequality, designed for two parties with two measurements, having two 
outcomes each. Not all entangled states violate the CHSH inequality \cite{Werner1989a}, and 
one may wonder whether the usage of measurements with more outcomes helps in
observing nonclassical behaviors. Is there a quantum state for which the 
infinite set of all possible two-outcome measurements does not lead to nonclassical 
effects, but some three-outcome measurements lead to a Bell inequality violation? 
This question has not been answered despite decades of research, arguably due to 
the complex structure of Bell correlations.  

There are, however, other nonclassical correlations in quantum mechanics
besides the violation of Bell inequalities. An important one is captured
by the notion of quantum steering~\cite{Uola2019a,Cavalcanti2016a}. This phenomenon goes
back to Schr\"odinger's observation that in the Einstein-Podolsky-Rosen
argument, one party (typically called Alice) can steer the state of the 
other party (called Bob) by making suitable measurements~\cite{Schroedinger1935a}. The modern formulation 
of this effect has been given by Wiseman and coworkers \cite{Wiseman2007a} 
and since then it was found to be connected to many subjects in quantum 
information processing. For instance, it has been shown that the 
measurements made by Alice have to be incompatible, implying that 
commuting measurements as in classical physics are not suitable~\cite{Uola2015a,Quintino2014a}. Furthermore, the theory of quantum steering has turned 
out to be useful to solve long standing open problems concerning Bell 
inequalities, e.g., the construction of states having a positive partial
transpose, but violating a Bell inequality~\cite{Moroder2014a,Vertesi2014a}.

\begin{figure}[!t]
\begin{center}
\begin{minipage}{0.35\textwidth}
\includegraphics[width=\textwidth]{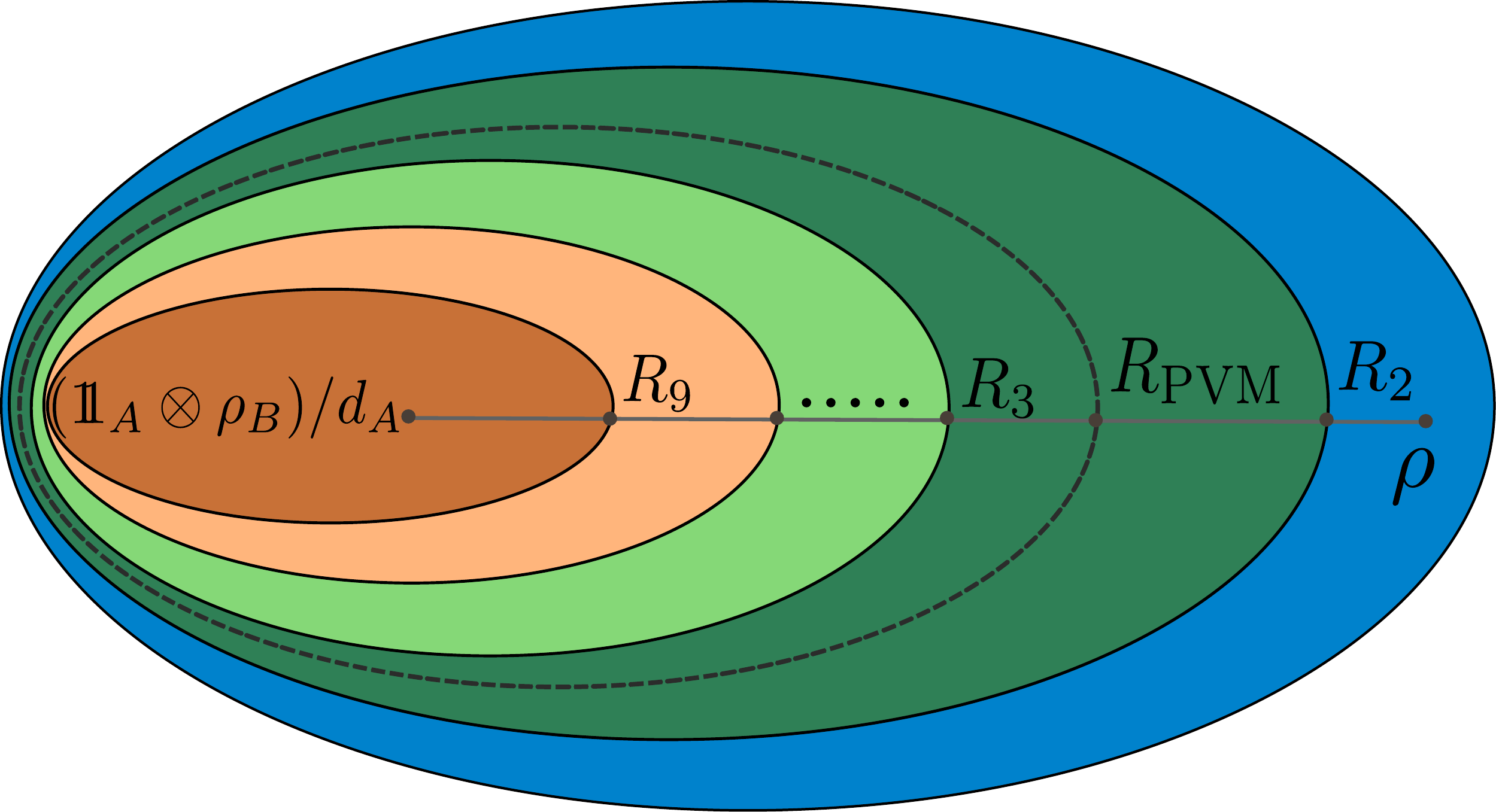}
\end{minipage}
\end{center}
\caption{The set of all bipartite quantum states (here for the case of dimenion $d=3$) 
can be divided into subsets, 
depending on how many outcomes measurements must have in 
order to show steering, and some states (dark brown, inner most) are not steerable at all.
For a given state $\rho$, the distance to these sets defines the
so-called critical radii $R_k$ or $R_{\mathrm{PVM}}$ for projective measurements. We prove that in general 
$R_3 > R_2$, demonstrating that some states require three-outcome
measurements for steering.}
\label{fig:layers}
\end{figure}  

The goal of this paper is twofold. First, we will show that for some 
quantum states a finite number of measurements of three outcomes can 
reveal quantum steering, while the infinite set of \emph{all} measurements with two 
outcomes cannot. This proves that the number of measurement outcomes
can be important to the question whether nonclassical effects can be 
observed or not. We note that in recent works it has been demonstrated 
that the correlations of certain multi-outcome measurements cannot be
explained by assuming that these measurements themselves have only two
effective outcomes~\cite{Kleinmann2016a,Kleinmann2017a,Hu2018a}. 
Thus this does not concern the fundamental limitation of the whole infinite set of two-outcome measurements as comparison to those with more outcomes in revealing quantum correlations.

Second, the methods developed in this paper allow one to advance the theory
of quantum steering in several directions. In particular, we derive
novel criteria for steerability and unsteerability, and present 
significantly improved local hidden variable models for so-called Werner 
states, which show that they do not violate any Bell inequality, even
if the most general measurements are considered~\cite{Barrett2002a}.

{\it Quantum steering.---}
Consider the situation where Alice and Bob share a bipartite 
quantum state $\rho$ and Alice performs a measurement (denoted 
by $x$) with $n$ outcomes. This is generally described by a collection 
of $n$ positive operators, $\{E_a^{(x)}\}_{a=1}^n$, $E_a^{(x)} \ge 0$, 
normalized by $\sum_{a=1}^{n} E_a^{(x)} = \I$, which form a so-called
positive operator valued measure (POVM). Bob's system is then found in 
the ensemble of conditional states $\{\rho_{a|x} = \Tr_A [\rho (E_a^{(x)} \otimes \I)]\}$. 
It has been noted early that by choosing different measurements, Alice can steer 
Bob's system to ensembles that are intuitively `incompatible' with each other, 
such as pure eigenstates of noncommutative observables, conflicting with our 
intuition of classical locality~\cite{Einstein1935a,Schroedinger1935a}. However, 
it was not until 2007 that this naive notion of `incompatibility' gained a 
precise definition.~\citet{Wiseman2007a} pointed out that incompatible 
ensembles in general mean that they cannot be derived from a single 
collection of states, called a local hidden state (LHS) ensemble. 
a LHS ensemble is simply a distribution $\mu$ on Bob's pure states 
$\ket{\lambda}$. The different 
ensembles $\{\Tr_A [\rho (E_a^{(x)} \otimes \I)]\}$ corresponding to 
different measurement choices $x$ can be derived from 
the single LHS ensemble $\mu$ if one can reach any conditional state a LHS
$\Tr_A [\rho (E_a^{(x)} \otimes \I)]$ from the states $\ket{\lambda}$
via classical postprocessing. That means that there are  probabilities 
$G_a^{(x)}(\lambda)$ such that
\begin{equation}
	\Tr_A[\rho (E_a^{(x)} \otimes \I)]= \int \! \d \mu (\lambda) G_a^{(x)}(\lambda) \ketbra{\lambda}{\lambda},
\end{equation}
where the integration is taken over Bob's pure states. If this is the case, 
one says that $\rho$ admits a LHS model, or in short, $\rho$ is \emph{unsteerable}. 
The postprocessing functions $G_a^{(x)}(\lambda)$ are called Alice's \emph{response 
functions}. Being probabilities, the response functions $G_a^{(x)}(\lambda)$ are 
constrained by $0 \le G_a^{(x)}(\lambda) \le 1$, $\sum_{a=1}^{n} G_a^{(x)}(\lambda)=1$. 
If such a LHS model does not exist, one says that 
$\rho$ is \emph{steerable}~\cite{Wiseman2007a}.

{\it The role of measurements.---}
Crucially for our purpose, Alice's steering abilities 
depend on the set of measurements $\M$ she can potentially 
make. This allows one to quantify how much steering the 
measurements of a class $\M$ reveal for a state $\rho$. 
Specifically, we define the steering \emph{critical radius} \red{of $\rho$ with respect to measurements in $\M$, denoted
$R_\M(\rho)$,} to be the maximum of the mixing parameter $\eta$ 
such that $\rho_\eta= \eta \rho + (1-\eta) (\I_A \otimes \rho_B)/d_A$ 
is unsteerable with measurements in $\M$,
\begin{equation}
R_\M(\rho)= \max \{\eta \ge 0: \mbox{$\rho_\eta$ is unsteerable w.r.t. $\M$}\}.
\label{eq:mixing}
\end{equation}
Here $\I_A$ denotes the identity operator acting on system $A$ 
and $\rho_B$ denotes the reduced state of system $B$, $\rho_B=\Tr_A (\rho)$. 
Geometrically $1-R_{\M} (\rho)$ measures the distance from $\rho$ to the 
surface separating steerable/unsteerable states (with measurements in $\M$) 
relatively to the noisy and unsteerable state $(\I_A \otimes \rho_B)/d_A$, 
see also Fig.~\ref{fig:layers}. \red{This special choice of the separable state is explained in~Ref.\cite[Section A]{sm}. There we also show how this definition stems from the critical radius defined in Ref.~\cite{Nguyen2018b}, originally measures certain geometrical object associated to a two-qubit system.} 
In a similar fashion, we 
 define $S(\rho)$ to be the maximum mixing parameter $\eta$ such that 
$\rho_\eta$ \red{is} separable, i.e., it can be written as a convex 
combination of product states~\cite{Werner1989a}.

{\it The structure of measurements.---}
The set of POVMs has a nested structure: measurements with 
$n$ outcomes are naturally a subset of that of measurements 
with $n+1$ outcomes.  Measurements with two outcomes, so-called 
dichotomic measurements, are the most elementary, and also 
among \red{the measurements that are most often performed} 
in experiments. Measurements whose effects $E_a$ are rank-$1$ 
projections will be referred to as projective measurements which are the standard measurements occurring 
in textbooks.

For $\M$ being the set of POVMs of $n$ outcomes, or projective 
measurements, we simply denote the critical radii by $R_n$, and 
$R_{\PVM}$, respectively. Since any POVM can be written as a mixture 
of POVMs with at most $d^2$ outcomes, measurements with $n > d^2_A$ outcomes
do not bring any more steerability to Alice~\cite{Barrett2002a,Adriano2005a}. 
So we can also denote $R_{\POVM}=R_{d^2_A}$. Because measurements with 
$n$ outcomes form a subset of that with $n+1$ outcomes, and projective 
measurements form a subset of measurements with $d_A$ outcomes, the critical 
radii organize in the following sequence
\begin{equation}
\begin{array}{cccccl}
 &  &  & & R_{\PVM} & \\
  & &  & &   \rotatebox[origin=c]{-90}{$\ge$} &\\
R_2 \! & \ge & \!  \cdots \! & \ge & R_{d_A} & \ge  \cdots   \ge R_{d^2_A} = R_{\POVM}, \\
\end{array}
\label{eq:R-hierarchy}
\end{equation}
which is valid for any state. Fig.~\ref{fig:layers} 
illustrates this sequence geometrically.

Although difficult to compute, already in their early 
paper,~\citet{Wiseman2007a} remarked that $R_{\PVM}$ can be computed for the Werner states and the isotropic states. More recently, it has been shown 
that $R_2$ can also be computed for arbitrary two-qubit
states~\cite{Jevtic2015a,Nguyen2016a,Nguyen2016b,Nguyen2018b}. 
Further, numerical evidences suggested that for two-qubit states, 
the chain in fact collapses to a single value $R_2=R_{\PVM}=R_3=R_4$~\cite{Nguyen2018b}.  

Here we report a practically closed formula for $R_2$ for the high-dimensional 
isotropic states and Werner states and show that $R_2 > R_{\PVM} \ge R_{\POVM}$ 
for systems other than qubits. This is in particular true for 
dimension $d=3$: $R_2 > R_3$ for the three-dimensional isotropic 
and Werner states. 
Since by replacing the infinite set of 
$3$-POVMs by a finite subset of measurements, one can 
approach $R_3$ (from above) as close as possible. So, there 
exists a finite set of measurements of three outcomes which 
gives a smaller critical radius than $R_2$. These three-outcome
measurements can then reveal nonclassicality, where all two-outcome
measurements cannot.

\begin{figure}[!t]
\begin{minipage}{0.235\textwidth}
\includegraphics[width=\textwidth]{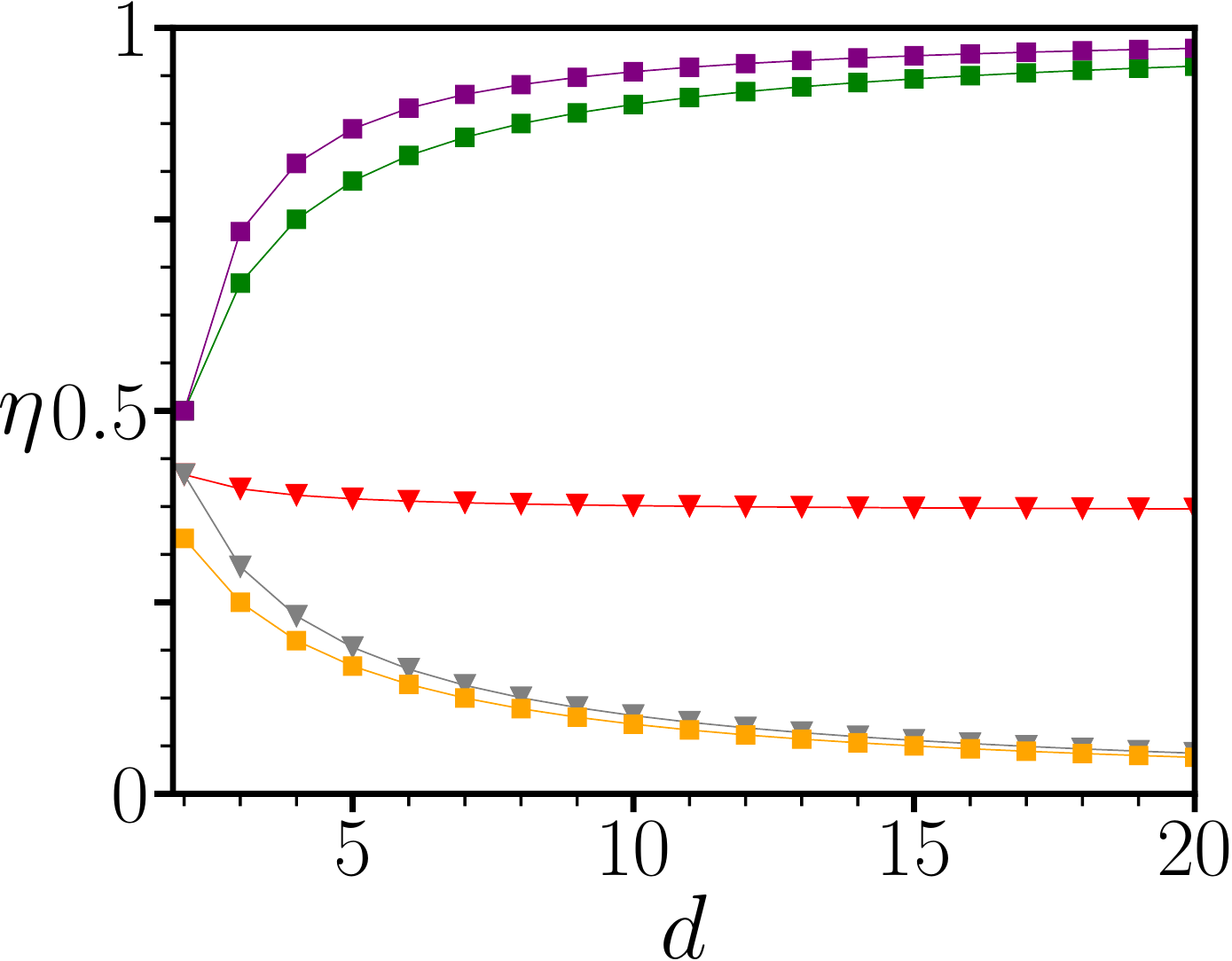}
\end{minipage}
\begin{minipage}{0.225\textwidth}
\includegraphics[width=\textwidth]{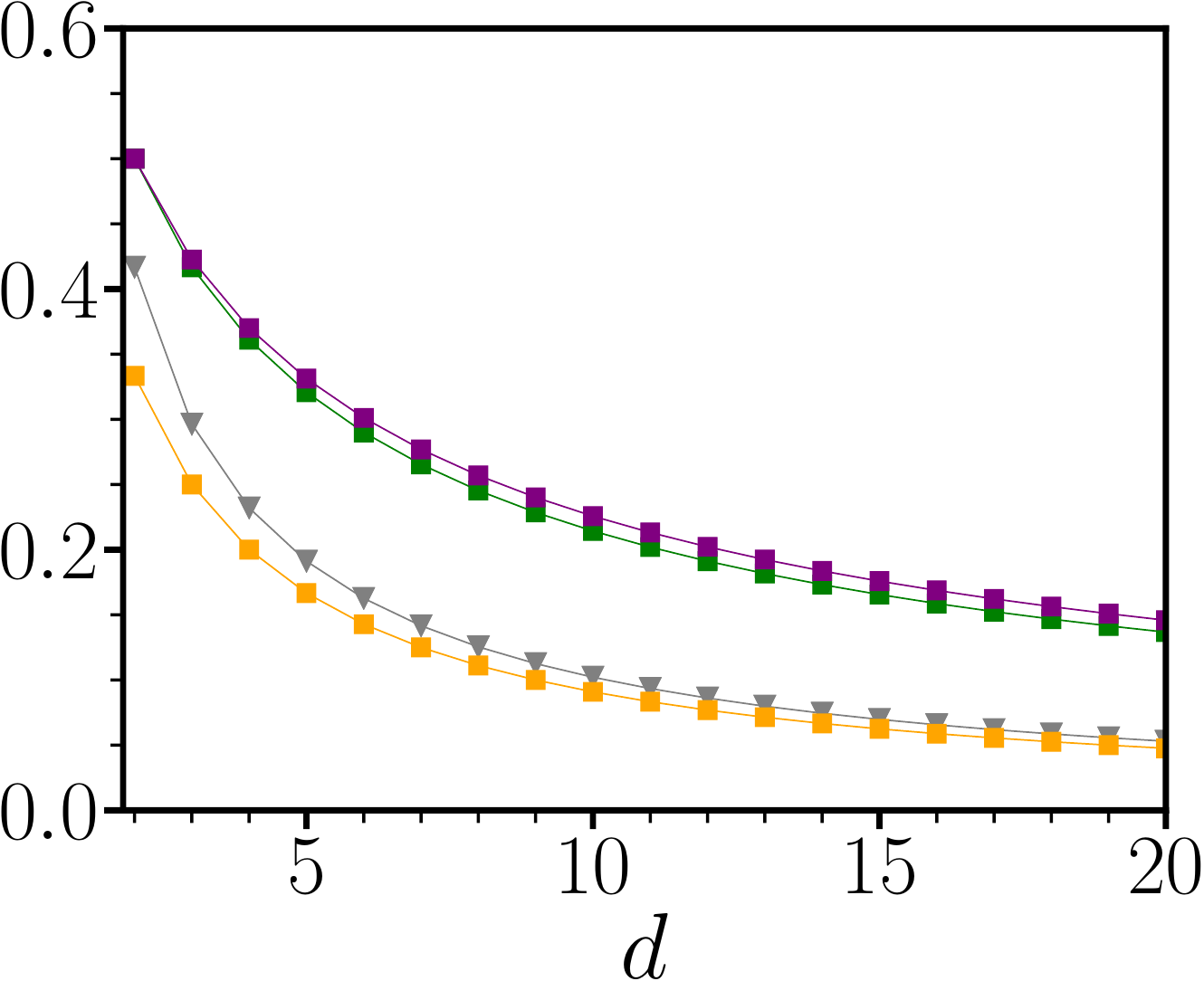}
\end{minipage}
\caption{Summary of the results on the steering critical radii
for Werner states (left) and isotropic states (right). From top
to bottom, we show the steering critical radii $R_2$ for
dichotomic measurements from Eqs.~(\ref{eq-werner-r2},
\ref{eq-iso-r2}) (violet), $R_{\PVM}$ for projective measurements 
from Ref.~\cite{Wiseman2007a} (green), a lower bound on $R_{\POVM}$
for Werner states from Eq.~(\ref{eq:etawb}) (red), lower bounds on $R_{\POVM}$
	from Ref.~\cite{Barrett2002a,Almeida2007a} (grey), and the separability limit $S$ (orange).}
	\label{fig:exact}
\end{figure}

{\it Werner states and the isotropic states.---}
Recall that the fully antisymmetric state of dimension 
$d \times d$ is defined by $W^d= {2 \pi_-}/{(d^2-d)}$,
where $\pi_-$ is the projection onto the antisymmetric 
subspace of $\CC^d \otimes \CC^d$, spanned by vectors
of the type $\ket{ij} - \ket{ji}$ \cite{Werner1989a}. 
The Werner state at mixing probability $\eta$ is then 
defined by mixing this projection with the white noise,  
$W_\eta^d= \eta W^d + (1-\eta) (\II/d) \otimes (\II/d)$.
This is in line with the notation introduced before
Eq.~(\ref{eq:mixing}), as we have $\Tr_A(W^d)= \II/d.$

By construction, the Werner states are symmetric under 
application of the same local unitary operation 
$U \in \mathrm{U} (d) $ on both parties, namely, 
$W_\eta^d = (U \otimes U) W_\eta^d (U^\dagger \otimes U^\dagger)$~\cite{Werner1989a}. 
It has been shown that  Werner states are separable if 
and only if \red{$\eta \le 1/(d+1)$}~\cite{Werner1989a}, which, 
can be written in the above notation as $S(W^d)=1/(d+1)$. 
Werner states are unsteerable with projective measurements 
if and only if \red{$\eta \le 1-1/d$}~\cite{Werner1989a, Wiseman2007a}, 
thus $R_{\PVM}(W^d)=1-1/d$. 

To define the isotropic states, one first considers 
the maximally entangled state on $\CC^d \otimes \CC^d$,
defined by $S^d= \ketbra{\phi_+}{\phi_+}$,
where $\ket{\phi_+}={1}/{\sqrt{d}} \sum_{k=1}^{d} \ket{k} \otimes \ket{k}$. 
The isotropic state at mixing probability $\eta$ is then 
$S^d_\eta= \eta S^d + (1- \eta) (\II/d) \otimes (\II/d)$. The isotropic state 
also has a symmetry under local  unitaries $U \in \mathrm{U} (d)$, as 
$S_\eta^d=  (U \otimes U^*) S_\eta^d ({U}^\dagger \otimes {(U^*)}^\dagger)$, 
where ${U^*}$ stands for the complex conjugate of $U$~\cite{Horodecki1999a}.
It is well-known that $S(S^d)=1/(d+1)$~\cite{Horodecki1999a}, 
and $R_{\PVM}(S^d)=(H_d-1)/(d-1)$, where $H_d=1+1/2+\cdots+1/d$~\cite{Wiseman2007a}.

{\it The uniform distribution as LHS ensemble.---} 
When writing down a LHS model as in Eq.~(\ref{eq:mixing}) for
Werner states or isotropic states, it is known~\cite{Wiseman2007a,Nguyen2018a} 
that one can restrict the attention to a probability distribution which is the uniform distribution according to the Haar measure,
denoted by $\omega$, over Bob's Bloch sphere. It is easily see from the argument 
given in Ref.~\cite{Nguyen2018a} that this remains true also 
if the measurements are limited to generalised ones of 
any fixed number of outcomes.

To proceed, we consider the set of conditional states Alice can 
simulate using this distribution $\omega$, which is given by \red{the convex set}
\begin{equation}
\K (\omega)= \left\{K = \int \! \d \omega (\lambda) g(\lambda) \ketbra{\lambda}{\lambda}: 0 \le g(\lambda) \le 1\right\}.
\end{equation} 
The set $\K (\omega)$ is known as the capacity of $\omega$~\cite{Nguyen2018a,Nguyen2018b}. 
In higher-dimensional spaces, $\K (\omega)$ has complicated structure and no
complete characterization of its geometry is known. However, we will see that 
even a partial information of $\K (\omega)$ will be sufficient to characterize 
quantum steering of Werner states and isotropic states.

{\it Dichotomic measurements.} 
Each dichotomic measurement is completely characterized by 
one of its two effects, say $M$, since the other is 
$\II - M$. It follows directly from the definition of 
quantum steering that Werner states and isotropic 
states are unsteerable if and only if the corresponding 
conditional state $\Tr_A[\rho (M \otimes \II)]$ is inside 
$\K(\omega)$ \red{(and so is $\Tr_A[\rho ((\II - M) \otimes \II)]$)} for all measurement effects $M$ on Alice's side.

Let us have a closer look at the set of measurement effects 
on Alice's side, $\{M: 0 \le M \le \II\}$. This is a convex set, 
of which the extreme points are precisely the projection operators.
These can be organized in hyperplanes corresponding to different 
ranks of the projections. It is then natural to introduce finer 
subsets of $2$-POVMs whose two effects are projections and the 
lower rank is $r$. Accordingly, we use $R_2^r(\rho)$ to denote 
the steering critical radius corresponding to this subset of 
measurements. We then have
\begin{equation}
R_2  = \min_{r=1,\ldots,\lfloor d/2 \rfloor} R_2^r, 
\label{eq:minimise_r}
\end{equation}
where $\lfloor d/2 \rfloor$ is the maximal integer not 
greater than $d/2$.

{\it Reducing the dimension and main result.---} 
The following observation is crucial to computing  $R_2^r$: 
For Werner states and isotropic states, a conditional state 
of Bob's system corresponding to a projection $P$ on Alice's 
side belongs to a special two-dimensional plane spanned by 
the projection itself and the identity operator, $\lspan\{P,\II\}$. 
This is easily verified by direct computation of the conditional states 
in these cases. Consequently, instead of 
considering the general capacity $\K(\omega)$, we can consider 
its cross-section with these two-dimensional subspaces and the 
original high-dimensional problem is now reduced to a 
two-dimensional one. Fortunately, in these two-dimensional 
spaces, the cross-section with $\K (\omega)$ can be computed 
exactly. The formulae are somewhat cumbersome, but can be 
explicitly given~\cite[Section B and C]{sm}. To find the critical 
radii $R_2^r$ of the fully antisymmetric state and the 
maximally entangled state, we simply identify the critical 
mixing probability threshold at which Bob's conditional states 
corresponding to a projection of rank $r$ is at the border 
of this cross-section; for the details, see \cite[Section~C and D]{sm}. 

The remaining step is the discrete minimization of 
$R_2^r$ with respect to the rank $r$ of the projection in Eq.~\eqref{eq:minimise_r}. 
We find that for both Werner states and isotropic states, 
$R_2^r$ is always minimal at $r=1$ for all dimensions 
$d \le 10^5$ and conjecture that this holds in 
general. In other words, among dichotomic measurements, 
those with a rank-$1$ effects are conjectured to be most 
useful for quantum steering. This eventually leads to 
the steering critical radius 
\begin{equation}
R_2 (W^d) = (d-1)^2 [ 1- (1-1/d)^{1/(d-1)}]
\label{eq-werner-r2}
\end{equation}
for Werner states, and 
\begin{equation}
R_2 (S^d) = 1-d^{-1/(d-1)}
\label{eq-iso-r2}
\end{equation}
for isotropic states. These critical radii are presented in 
Fig.~\ref{fig:exact} together with other known thresholds 
for these two families of states.

As an example, for the system of two qutrits, $d=3$, we find for the Werner 
state $R_2 (W^3)=4(1-\sqrt{2/3})\approx 0.734$, which is strictly larger 
than $R_{\PVM}(W^3)=2/3 \approx 0.667$, and  for the isotropic state 
$R_2(S^3)=1-1/\sqrt{3}\approx 0.423$, which is also strictly larger 
than $R_{\PVM}(S^3)=5/12 \approx 0.417$. 

{\it Steering with arbitrary POVMs.---}
As long as quantum steerability is concerned, it follows 
from Ref.~\cite{Barrett2002a} that without loss of 
generality, one can assume that Alice's measurements 
consist of $d^2$  rank-$1$ effects, $E=(E_1,E_2,\cdots,E_{d^2})$ 
with $E_a = \alpha_a P_a$, where $P_a$ are rank-$1$ projections, 
$0 \le \alpha_a \le 1$, and $\sum_{a=1}^{d^2} \alpha_a = d$.
Let us consider the Werner state $W_\eta^d$.  For outcome $a$ of 
Alice's measurement, Bob's system is steered to 
$\Tr_A(W_\eta^d E_a \otimes \II) = \alpha_a \Tr_A [W^d_\eta P_a \otimes \I]$.
One sees that apart from the multiplication factor $\alpha_a$, the 
conditional states are essentially that of $n=d^2$ dichotomic 
measurements $(P_a,\I-P_a)$. But even if the state is unsteerable 
with dichotomic measurements and the explicit response functions 
are given, it is not possible to directly combine them to form a response 
function for the general POVM $E$, which requires the normalization 
for the response function as probabilities, $\sum_{a=1}^{d^2} G_a(\lambda) =1$. 
To achieve the normalization, one has to soften the response functions 
for the dichotomic measurements in a suitable way. Barrett was the 
first who used this idea to construct \red{a local hidden variable model with POVMs 
for certain entangled Werner states, which later turns out to be} a LHS model~\cite{Barrett2002a,Quintino2015a}. As it 
turns out, his construction is in fact most suitable 
when the two parties are correlated, such as when 
they share an isotropic state. For the Werner states, 
the two parties are however anticorrelated. 
We therefore propose the following response function 
for the Werner state, 
\begin{align}
G_a(\lambda)&=  \alpha_a \bra{\lambda}\frac{\II-P_a}{d-1}\ket{\lambda}  \Theta (1/d-\bra{\lambda}P_a\ket{\lambda}) \\
& \!\!\!\!\!\!\!\! + \frac{\alpha_a}{d} \left[1-\sum_{b=1}^{n} \alpha_b \bra{\lambda} \frac{\II-P_b}{d-1} \ket{\lambda} \Theta (1/d-\bra{\lambda}P_b \ket{\lambda}) \right].
\nonumber
\end{align} 
The physical intuition for this response function and detailed calculation are discussed in~\cite[Section E]{sm}. 
With this, direct computation gives 
\begin{equation}
R_{\POVM}(W^d) \ge \frac{1+(d-1)^{d+1}d^{-d}}{d+1}.
	\label{eq:etawb}
\end{equation}
Fig.~\ref{fig:exact} shows that this significantly improves the bound given 
by the original Barrett construction, in particular it remains finite as $d$ tends to infinity,
$\lim_{d \to \infty} R_{\POVM}(W^d) = 1/e$, where $e$ is the Euler's 
natural constant. Note that \red{constructing a LHS model for all POVMs for certain Werner state, our results also imply that the respective Werner states do not violate any Bell
inequality}.

{\it Steering criteria for general states.---}
We now show that our methods 
can be used 
to analyse steerability of generic high-dimensional states, where Bob's 
reduced state is of full rank. 
Because steerability is 
invariant under local filtering on Bob's side~\cite{Uola2014a,Quintino2015a,Rodrigo2015a}, we can assume Bob's reduced state to be maximally 
mixed (by applying an appropriate filtering). 

Then, one use the fact that the steerability from Alice to Bob 
is non-increasing under local channels on Alice's side~\cite{Baker2018a}. 
Given two states $\rho$ and $\tau$, each with Bob's reduced state 
maximally mixed, \red{we define 
$D(\rho,\tau)$ to be the maximum value of $\eta$ such that $\rho_{\eta}$ in Eq.~(\ref{eq:mixing}) can still be obtained from a certain local channel $\mathcal{E}$ on Alice's side acting on $\tau$, namely $\rho_\eta = (\mathcal{E} \otimes \mathcal{I}) [\tau]$, 
where $\mathcal{I}$ is the identity 
channel.} 
Slightly extending the result of~\cite{Baker2018a}, it directly follows that 
given an unsteerable state $\tau$, i.e., $R_n (\tau) \ge 1$, then
\begin{equation} 
R_{n} (\rho) \ge D(\rho,\tau).
\label{eq:lower_bound}
\end{equation}
Given $\tau$, the computation of $D(\rho,\tau)$ is a standard 
optimization over the \red{suitable} channel $\mathcal{E}$, which can be done
using semidefinite programming~\cite{Boyd2004a}. By 
choosing $\tau$ to be an unsteerable Werner state, or 
an unsteerable isotropic state, Eq.~\eqref{eq:lower_bound} 
gives a \emph{lower} bound for $R_{n} (\rho)$ and consequently
a way to prove the unsteerability of a generic high-dimensional 
state.
Interestingly, one can also turn the logic of Eq.~\eqref{eq:lower_bound} 
around and prove steerability. 
In this case, one chooses $\rho$ to be 
a state of which $R_n(\rho)$ is known
then $D(\rho,\tau)>R_{n} (\rho)$ implies 
that $R_n (\tau)<1$, which proves the steerability of $\tau$.

Another way to prove steerability is to average the state by random unitary such that it results in a Werner state or a isotropic state~\cite{Werner1989a,Horodecki1999a}.
During this process, the critical radius cannot decrease,   we thus find
\begin{equation}
R_n(\rho) \le  \min\left\{\frac{(d+1)R_n (W^d)}{1-d F_W},\frac{(d^2-1)R_n(S^d)}{d^2-F_S-1}\right\},
\end{equation}
where $F_S=\Tr(S^d \rho)$ and $F_W=\Tr(F^d \rho)$, with $F^d$ being the 
swap operator between two systems of dimension $d$. Such an upper bound 
allows one to prove the steerability of the state.

{\it Conclusion.---}
We showed that the number of outcomes of measurements is essential 
for their ability to reveal nonclassicality.
While we concentrated on quantum steering as a prototype quantum correlation, we conjecture that this also holds for Bell correlation. 
That is, there are states which admit a local hidden variable model for \emph{all} dichotomic measurements, nevertheless violating certain Bell inequalities for finite measurement settings with sufficient number of outcomes. 

An immediate application 
and extension of this phenomenon is the development of methods to \red{assess the 
quality} of measurements in experiments. In fact, \red{it can} be anticipated
that the results obtained here can lead to novel protocols for the 
self-testing of measurements in experiments. Furthermore, our results
may be used to characterize the resources needed for the simulation
of measurements, as there are situations where all dichotomic measurements 
can easily be simulated, while three-outcome measurements cannot. 

Moreover, we provided novel criteria for the steerability and unsteerability
of general quantum states. Especially the presented LHS model for Werner states 
improves the known models drastically. These results will be useful for the 
applications of steering in information processing, such as quantum key 
distribution in asymmetric scenarios~\cite{Branciard2012a}, or the characterization
of joint measureability~\cite{Uola2019a}.

\begin{acknowledgments}
We thank Travis Baker, S\'ebastien Designolle, 
Matthias Kleinmann, M. Toan Nguyen, Jiangwei Shang, and Roope Uola 
for inspiring discussions. This work was supported by the Deutsche Forschungsgemeinschaft (DFG, German
Research Foundation - 447948357) and the ERC (Consolidator Grant 683107/TempoQ). 
\end{acknowledgments}



\bibliography{quantum-steering}

\onecolumngrid
\newpage


\section*{Supplementary material}

# Supplementary Materials: Some quantum measurements with three outcomes can reveal nonclassicality where all two-outcome measurements fail to do so

H. Chau Nguyen[*] and Otfried Gühne[†]

*Naturwissenschaftlich-Technische Fakultät, Universität Siegen, Walter-Flex-Straße 3, 57068 Siegen, Germany*

(Dated: December 10, 2020)

## Appendix A: More on the geometry of the critical radius

In this section, we discuss in more detail the definition of the critical radius, Eq. (2) in the maintext. In particular, we clarify the geometrical motivation of the choice of the separable noisy state $(\mathbb{1}_A \otimes \rho_B)/d$.

Let us first remark that the critical radius in Ref. [1] was originally defined differently (as a measure of certain geometrical aspect of the state *rho* itself); but one can show that it is equivalent to definition Eq. (2) in the maintext with that particular choice of the separable noisy state $(\mathbb{1}_A \otimes \rho_B)/d$. Within our current context, we can also motivate this choice as follows. Let us consider the attempt to define the critical radius as

$$R_\mathcal{M}(\rho) = \max\{\eta \geq 0 : \rho_\eta \text{ is unsteerable w.r.t. } \mathcal{M}\}. \tag{A1}$$

with the more general form of $\rho_\eta = \eta \rho + (1-\eta)(\tau_A \otimes \tau_B)$.

Consider the family of sets of steering outcomes $\{\mathrm{Tr}_A[\rho_\eta(M \otimes \mathbb{1})] : 0 \leq M \leq \mathbb{1}\}$ with $\rho_\eta = \eta\rho + (1-\eta)\tau_A \otimes \tau_B$ for $0 \leq \eta \leq 1$. We then see that if $\tau_B = \rho_B$, this family shares the same common point $\rho_B = \mathrm{Tr}_A(\rho_\eta)$ for all $\eta$.

The choice $\tau_A = \mathbb{1}_A/d$ is a bit subtler. As we also mentioned in the main text, the extreme points of the set $\{M : 0 \leq M \leq \mathbb{1}_A\}$ are projections, which organise in planes of operators with integer traces. Therefore the extreme points of the set $\{\mathrm{Tr}_A[(M \otimes \mathbb{1}_B)\rho_\eta] : 0 \leq M \leq \mathbb{1}_A\}$ for certain $\eta$ also organise in a set of parallel planes (linear images of the planes of operators of integer traces). If $\tau_A = \mathbb{1}_A/d$, this set of parallel planes remain constant for all the sets $\{\mathrm{Tr}_A[(M \otimes \mathbb{1}_B)\rho_\eta] : 0 \leq M \leq \mathbb{1}_A\}$ as $\eta$ varies.

These two geometrical features have imprints on the computational aspect of the critical radius. We are to demonstrate this implication below and link the definition (A1) to the suggestion in Ref. [1, Supplementary X].

The key step in making the definition (A1) in a computational form, at least in principle, is to use the results of Ref. [5, Theorem 1], which states that $\rho_\eta$ is unsteerable if and only if for some choice of a distribution $\mu$ over Bob's Bloch sphere (i.e., a choice of LHS ensemble), one has

$$\int \mathrm{d}\mu(\sigma) \max_i \mathrm{Tr}(Z_i \sigma) \geq \sum_{i=1}^n \mathrm{Tr}[\rho_\eta(E_i \otimes Z_i)] \tag{A2}$$

for all measurements $E = (E_1, E_2, \ldots, E_n)$ on Alice's side and arbitrary choice of observables $Z = (Z_1, Z_2, \ldots, Z_n)$ on Bob's side. Thus the critical radius can be formally written as

$$\begin{aligned}
\max_{\eta,\mu} &\quad \eta \\
\text{s.t.} &\quad \eta \geq 0, \\
&\quad \int \mathrm{d}\mu(\sigma) \max_i \mathrm{Tr}(Z_i \sigma) \geq \sum_{i=1}^n \mathrm{Tr}[\rho_\eta(E_i \otimes Z_i)] \text{ for all } E, Z, \\
&\quad \int \mathrm{d}\mu(\sigma) = \mathrm{Tr}_A(\rho_\eta).
\end{aligned} \tag{A3}$$

By the choosing $\tau_B = \rho_B$, the second constraint is independent of $\eta$. As a result, one can then rewrite the optimisation problem as

$$\begin{aligned}
\max_{\eta,\mu} &\quad \eta \\
\text{s.t.} &\quad \eta \geq 0, \\
&\quad \int \mathrm{d}\mu(\sigma)\{\max_i \mathrm{Tr}(Z_i \sigma) - \sum_{i=1}^n \mathrm{Tr}(E_i \tau_A) \mathrm{Tr}(Z_i \sigma)\} \geq \eta \sum_{i=1}^n \mathrm{Tr}[(\rho - \tau_A \otimes \rho_B)(E_i \otimes Z_i)] \text{ for all } E, Z, \\
&\quad \int \mathrm{d}\mu(\sigma) = \rho_B.
\end{aligned} \tag{A4}$$

---

[*] chau.nguyen@uni-siegen.de
[†] otfried.guehne@uni-siegen.de

The importance of the choice $\tau_A = \mathbb{1}_A/d$ becomes apparent when we assume that the measurement $E$ is projective. Indeed, if $\tau_A = \mathbb{1}_A/d$, and the measurement is rank-1 projective, then $\text{Tr}(E_i \tau_A) = 1/d$ for all $i$. Thus one can further rewrite the optimisation problem as

$$\begin{aligned}
\max_{\eta,\mu} \quad & \eta \\
\text{s.t.} \quad & \eta \geq 0 \\
& \int d\mu(\sigma)\{\max_i \text{Tr}(Z_i \sigma) - 1/d \sum_{i=1}^n \text{Tr}(Z_i, \sigma)\} \geq \eta \max_E \{\sum_{i=1}^n \text{Tr}[(\rho - \tau_A \otimes \rho_B)(E_i \otimes Z_i)]\} \text{ for all } Z \\
& \int d\mu(\sigma) = \rho_B
\end{aligned} \quad (A5)$$

The seemingly minor difference between (A5) and (A4) is in fact very important. One way to see it is to consider approximating the Bloch sphere by a set of discrete points $\sigma$. The effect of this discretisation is that in the second constraint of the optimisation problems (A4) and (A5), one only needs to consider a finite number of choices for $Z$; we refer the readers to Ref. [1] for a more detailed discussion of this discretisation procedure. Then for the problem (A5), upon solving a finite number of optimisation over measurements $E$ corresponding to different choices of $Z$ in the second constraint, the computation of the critical radius as the optimal $\eta$ is then a linear program (with finite number of constraints). On the other hand, the problem (A4) cannot be brought into a linear program (with finite number of constraints) despite the finite number of choices of $Z$.

From Eq. (A4), one also see that the definition of the critical radius directly link to that suggested in Ref. [5, Supplementary X]. Indeed, the problem (A4) can be rewritten as

$$R^{-1}[\rho] = \inf_\mu \sup_{E,Z} \frac{\sum_{i=1}^n \text{Tr}[(\rho - \tau_A \otimes \rho_B)(E_i \otimes Z_i)]}{\int d\mu(\sigma)\{\max_i \text{Tr}(Z_i \sigma) - \sum_{i=1}^n \text{Tr}(E_i \tau_A) \text{Tr}(Z_i \sigma)\}}, \quad (A6)$$

subject to $\int d\mu(\sigma)\sigma = \rho_B$. In doing so, we note that $\int d\mu(\sigma)\{\max_i \text{Tr}(Z_i \sigma) - \sum_{i=1}^n \text{Tr}(E_i \tau_A) \text{Tr}(Z_i, \sigma)\} \geq 0$. It is clear that Eq. (A6) is the same as the suggestion for an extension of critical radius as in Ref. [5, Supplementary X] with $\tau_A = \mathbb{1}_A/d_A$.

While we have presented the motivation for the choice of the separable state $(\mathbb{1}_A \otimes \rho_B)/d$ in the definition of the critical radius, this does not rule out the possible choices. Further exploration of these possibilities can be an interesting research direction.

## Appendix B: Integration over the high dimensional Bloch sphere

We will frequently have to work with integrals over the high dimensional Bloch sphere (i.e., the set of pure states). Here we describe how that can be done, following Refs. [2, 3] with small modifications.

Specifically, we work with the Hilbert space of dimension $d$. Let $Q$ be a projection of rank $k$, we are interested in the following integration

$$a_n(k,t) = \int d\omega(\lambda) \langle \lambda | Q | \lambda \rangle^n \Theta(\langle \lambda | Q | \lambda \rangle - t), \quad (B1)$$

where $\Theta$ is Heaviside's step function and $\omega$ denotes the Haar measure over the pure states. Note that although the projection $Q$ appears in the integral on the right-hand side, we will see that the left-hand side only depends on its rank $k$, which justifies the notation $a_n(k,t)$.

We choose the basis $\{|i\rangle\}_{i=1}^d$ such that $Q = \sum_{i=1}^k |i\rangle\langle i|$. The pure state can be written as $|\lambda\rangle = \sum_{i=1}^d r_i e^{i\theta_i} |i\rangle$. The Haar measure thus can be formally written as

$$d\omega(\lambda) = \frac{1}{Z} \prod_{i=1}^d r_i dr_i d\theta_i \delta(\sum_{i=1}^d r_i^2 - 1). \quad (B2)$$

The range of $r_i$ is $[0, +\infty)$ and the range of $\theta_i$ is $[0, 2\pi)$. The normalisation factor $Z$ can be found by

$$Z = \prod_{i=1}^d \int_0^{+\infty} r_i dr_i \int_0^{2\pi} d\theta_i \delta(\sum_{i=1}^d r_i^2 - 1). \quad (B3)$$

Now note that the integrands in (B1) and (B3) do not depend on the phase $\theta_i$, thus the integration over the phase $\theta_i$ can be carried out directly. Moreover, the integrals over $r_i$ can be simplified by changing the variable $u_i = r_i^2$. Eventually, we obtain

$$a_n(k,t) = \frac{I_n(Q,t)}{I_0(Q,0)}, \quad (B4)$$

with

$$I_n(Q,t) = \int du \delta(1 - \sum_{i=1}^{d} u_i)\Theta(\sum_{i=1}^{k} u_i - t)(\sum_{i=1}^{k} u_i)^n, \tag{B5}$$

where $du = du_1 du_2 \ldots du_d$ and the integral is taken over the whole range $[0, +\infty)$ of $u_i$.

Let

$$s_p(\xi) = \int dx_1 dx_2 \ldots dx_p \delta(\xi - x_1 - x_2 - \ldots - x_p). \tag{B6}$$

Then by rescaling the integral variable, one can easily show that

$$s_p(\xi) = s_p(1)\xi^{p-1}. \tag{B7}$$

Note that $s_p(1)$ is simply the area of the $p-1$ probability simplex, which still carries a $\delta$-function.

With this notation, we then can integrate out $u_{k+1}, u_{k+2}, \ldots, u_d$ in (B5) to get

$$I_n(Q,t) = s_{d-k}(1) \int du_1 \ldots du_k \Theta(\sum_{i=1}^{k} u_i - t)(1 - \sum_{i=1}^{k} u_i)^{d-k-1}(\sum_{i=1}^{k} u_i)^n. \tag{B8}$$

To carry out this integral, we write

$$I_n(Q,t) = s_{d-k}(1) \int du_1 \ldots du_k \Theta(\sum_{i=1}^{k} u_i - t)(1 - \sum_{i=1}^{k} u_i)^{d-k-1}(\sum_{i=1}^{k} u_i)^n \int_0^1 dx \delta(x - \sum_{i=1}^{k} u_i). \tag{B9}$$

Upon changing the integral order, we have

$$I_n(Q,t) = s_{d-k}(1) \int_0^1 dx \int du_1 \ldots du_k \Theta(x - t)(1 - x)^{d-k-1} x^n \delta(x - u_1 - u_2 - \cdots - u_k) \tag{B10}$$

$$= s_{d-k}(1) s_k(1) \int_t^1 dx (1-x)^{d-k-1} x^{k-1+n} \tag{B11}$$

$$= s_{d-k}(1) s_k(1) \beta(1-t, d-k, k+n), \tag{B12}$$

where $\beta(z, a, b) = \int_0^z d\xi \xi^{a-1}(1-\xi)^{b-1}$ is Euler's incomplete $\beta$-function. So

$$a_n(k,t) = \frac{\beta(1-t, d-k, k+n)}{\beta(d-k, k)}, \tag{B13}$$

where $\beta(a,b) = \beta(1, a, b)$ is Euler's complete $\beta$-function. As we remarked in the paragraph following (B1), $a_n(k,t)$ only depends on the rank $k$ of the projection $Q$.

### Appendix C: The canonical cross-sections of the capacity of the uniform distribution

Generally, it has been shown [4, 5] that the extreme points of $\mathcal{K}(\omega)$ are of the form

$$K(Z) = \int d\omega(\lambda)\Theta(\langle\lambda| Z |\lambda\rangle)|\lambda\rangle\langle\lambda|, \tag{C1}$$

with varying operator $Z$. In particular, let us consider a special family of these extreme points where $Z = Q - t\mathbb{1}$, where $Q$ is a (fixed) projection of rank $k$ and varying $t$,

$$K(Q,t) = \int d\omega(\lambda)\Theta(\langle\lambda| Q |\lambda\rangle - t)|\lambda\rangle\langle\lambda|. \tag{C2}$$

Let us now show that $K(Q,t)$ is in the span of $\{\mathbb{1}, Q\}$. While this can be done directly by inspection, a more elegant argument makes use of the concepts of von Neumann algebras [6]. Since $Q$ is a projection, the span of $\{\mathbb{1}, Q\}$ is also the von Neumann algebra generated by $\mathbb{1}_B$ and $Q$. To show that $K(Q,t)$ is in the algebra, we show that it

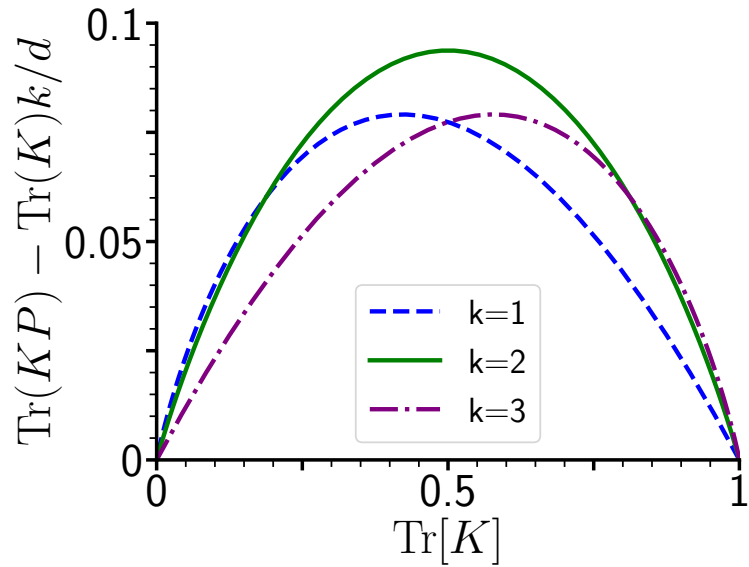

FIG. 1. The boundary of canonical cross-sections of the capacity of the uniform distribution specified in the plane spanned by $\{\mathbb{1}, Q\}$, with $\mathrm{Tr}(Q) = k$ for dimension $d = 4$. The operators $K$ on the boundary are specified by two orthogonal coordinates $\mathrm{Tr}(K)$ and $\mathrm{Tr}[K(P - \mathbb{1}k/d)] = \mathrm{Tr}(KP) - \mathrm{Tr}(K)k/d$.

commutes with all unitaries in the commutant of the span of $\{\mathbb{1}, Q\}$ [6]. That is, let $U$ be an unitary operator that commutes with $Q$, we want to show that $U$ also commute with $K(Q, t)$. Indeed,

$$UK(Q,t)U^\dagger = \int d\omega(\lambda) \Theta(\langle \lambda | Q | \lambda \rangle - t) U |\lambda\rangle\langle\lambda| U^\dagger. \tag{C3}$$

Upon transforming $|\lambda'\rangle = U |\lambda\rangle$ and noting that the Haar measure is invariant under this transformation, and that $\langle \lambda' | UQU^\dagger | \lambda' \rangle = \langle \lambda' | Q | \lambda' \rangle$) since $U$ commutes with $Q$, we obtain an identical formula as equation (C2) for $K(Q, t)$.

Being in the span of $\{\mathbb{1}, Q\}$, $K(Q, t)$ is characterised by two parameters $\mathrm{Tr}[K(Q,t)] = a_0(k, t)$ and $\mathrm{Tr}[QK(Q,t)] = a_1(k, t)$, with

$$a_0(k, t) = \frac{\beta(1-t, d-k, k)}{\beta(d-k, k)}, \tag{C4}$$

$$a_1(k, t) = \frac{\beta(1-t, d-k, k+1)}{\beta(d-k, k)}, \tag{C5}$$

as defined in Eq. (B1) and Eq. (B13).

As $t$ varying from 0 to 1, $K(Q, t)$ draws a curve starting at $\mathbb{1}$ and ending at 0 in the plane spanned by $\{\mathbb{1}, Q\}$. As a consequence, this forms a half of the boundary of the cross-section of $\mathcal{K}(\omega)$ in this plane; see Fig. 1. The other half of boundary of the cross-section is formed by $K(\mathbb{1} - Q, t)$ for $t$ varying from 0 to 1.

## Appendix D: Critical radii for dichotomic measurements

### 1. Werner states

Recall that the fully antisymmetric state of dimension $d \times d$ is defined by

$$W^d = \frac{2\pi_-}{d(d-1)}, \tag{D1}$$

where $\pi_-$ is the projection onto the antisymmetric subspace of $\mathbb{C}^d \otimes \mathbb{C}^d$. The Werner state acting in dimension $d$ is obtained as a convex combination of the fully antisymmetric state $W^d$ with the maximally mixed state,

$$W^d_\eta = \eta W^d + (1-\eta)\frac{\mathbb{1}}{d} \otimes \frac{\mathbb{1}}{d}, \tag{D2}$$

and $\eta$ is referred to as the mixing parameter.

Suppose Alice makes a dichotomic measurement $E = (P, \mathbb{1} - P)$, where $P$ is a projection of rank $r$. For the outcome $P$, Bob's system is steered to

$$\mathrm{Tr}_A(W_\eta^d P \otimes \mathbb{1}) = \eta \frac{\mathbb{1} - P}{d(d-1)} + \left[\eta \frac{r-1}{d-1} + (1-\eta)\frac{r}{d}\right] \frac{\mathbb{1}}{d}, \tag{D3}$$

where $r = \mathrm{rank}(P)$.

Observe that this steering outcome belongs to the plane spanned by $\mathbb{1}$ and $\mathbb{1} - P$. We consider the cross-section of the capacity of the uniform distribution $\mathcal{K}(\omega)$ in the corresponding plane, i.e., $K(\mathbb{1} - P, t)$, with the border described by equation (C5). We are interested in whether the conditional state (D3) is inside this cross-section. The condition for this to happen can be easily derived by identifying the critical value $\eta_c$ for the mixing parameter such that the conditional state (D3) is on the border of the capacity (C5), which is given by

$$a_0(d-r, t_c) = \frac{r}{d}, \tag{D4}$$

$$a_1(d-r, t_c) = \eta_c \frac{d-r}{d(d-1)} + \left[x_c \frac{r-1}{d-1} + (1-\eta_c)\frac{r}{d}\right]\left(1 - \frac{r}{d}\right). \tag{D5}$$

Solving $t_c$ from equation (D4), one can compute $\eta_c$ from equation (D5). Recall that $\eta_c$ is in fact precisely the definition of the critical radii, $R_2^r(W^d) = \eta_c$.

One can derive a more explicit formula for $R_2^r(W^d)$. Indeed, from equation (D5), we find

$$\eta_c = \frac{d^2(d-1)}{r(d-r)}\left[a_1(d-r, t_c) + \frac{r^2}{d^2} - \frac{r}{d}\right]. \tag{D6}$$

Upon using the definition of $a_n(k, t)$ in equation (B13), the recursive relation for the incomplete $\beta$-function [7, page 263],

$$\beta(z, a, b+1) = \frac{b}{a+b}\beta(z, a, b) + \frac{1}{a+b}z^a(1-z)^b, \tag{D7}$$

and the definition of the complete $\beta$-function in terms of the $\Gamma$-function [7, page 259],

$$\beta(a, b) = \frac{\Gamma(a)\Gamma(b)}{\Gamma(a+b)}, \tag{D8}$$

one arrives at

$$R_2^r(W^d) = \frac{(d-1)\Gamma(d+1)}{\Gamma(r+1)\Gamma(d-r+1)}(1-t_c)^r t_c^{d-r}. \tag{D9}$$

Although not given in a closed form for arbitrary $r$, $R_2^r(W^d)$ can be easily computed in a computer. For all $d \leq 10^5$, we compute $R_2^r(W^d)$ and find that it is always minimised at $r = 1$. Thus in all these cases we can identify $R_2$ with $R_2^1$. For $r = 1$, equation (D4) can be solved explicitly for $t_c$, and we arrive at

$$R_2(W^d) = (d-1)^2[1 - (1-1/d)^{1/(d-1)}]. \tag{D10}$$

### 2. Isotropic states

Recall that the maximally entangled state on $\mathbb{C}^d \otimes \mathbb{C}^d$ is defined by

$$S^d = |\psi_+\rangle\langle\psi_+|, \tag{D11}$$

where $|\psi_+\rangle = \frac{1}{\sqrt{d}}\sum_{k=1}^d |k\rangle \otimes |k\rangle$ for certain basis $|k\rangle$. The isotropic state is then defined by

$$S_\eta^d = \eta S^d + (1-\eta)\frac{\mathbb{1}}{d} \otimes \frac{\mathbb{1}}{d}. \tag{D12}$$

The computation of $R_2^r(S^d)$ follows similar steps as that for the Werner state. There is a remarkable difference, though. For the isotropic state, the steering outcome at Bob's side corresponding to the projection outcome $P$ at Alice's side,

$$\text{Tr}_A[S_\eta^d(P \otimes \mathbb{1})] = \frac{r}{d}\left[\eta\frac{\bar{P}}{r} + (1-\eta)\frac{\mathbb{1}}{d}\right], \tag{D13}$$

belongs to the canonical cross-section of the capacity of the uniform distribution indicated by $K(\bar{P}, t)$. Here $\bar{P}$ denotes the complex conjugate of $P$. Thus here we need to consider the cross-section of $\mathcal{K}(\omega)$ with the plan spanned by $\bar{P}$ and $\mathbb{1}$, in contrast to the case for the Werner states.

Following the same steps in Section D 1, we proceed by identifying the critical value $\eta_c$ for the mixing parameter such that the conditional state (D13) is on the border of the capacity (C5), which is given by

$$a_0(r, t_c) = \frac{r}{d}, \tag{D14}$$

$$a_1(r, t_c) = \eta_c \frac{r}{d} + (1-\eta_c)\frac{r^2}{d^2}. \tag{D15}$$

Solving $t_c$ from equation (D14), one can compute $\eta_c$ from equation (D15). Again, $\eta_c$ is in fact precisely the definition of the critical radii, $R_2^r(S^d) = \eta_c$.

An explicit formula for $R_2^r(S^d)$ can also be derived. From equation (D15), we find

$$\eta_c = \frac{a_1(r, t_c) - r^2/d^2}{r/d(1 - r/d)}. \tag{D16}$$

Then using the definition of $a_n(k, t)$ in equation (B13), the recursive relation (D7) and the relation between $\beta$-function and $\Gamma$-function (D8), one obtains

$$R_2^r(S^d) = \frac{\Gamma(d+1)}{\Gamma(d-r+1)\Gamma(r+1)}(1-t_c)^{d-r}t_c^r. \tag{D17}$$

Note the difference with the equation (D9) for the Werner state. For all $d \leq 10^5$, we again find that $R_2^r$ is minimised at $r = 1$. We thus have for all $d \leq 10^5$,

$$R_2(S^d) = 1 - d^{-1/(d-1)}. \tag{D18}$$

### Appendix E: New local hidden state model for the Werner states with generalised measurements

In the following, we present the details of the derivation of the bound

$$R_{\text{POVM}}(W^d) \geq \frac{1 + (d-1)^{d+1}d^{-d}}{d+1}. \tag{E1}$$

This bound is the critical mixing pramameter $\eta_c$ such that

$$\text{Tr}(W_{\eta_c}^d E_a \otimes \mathbb{1}) = \int d\omega(\lambda) G_a(\lambda)|\lambda\rangle\langle\lambda|, \tag{E2}$$

with the response function

$$G_a(\lambda) = \alpha_a \langle\lambda|\frac{\mathbb{1} - P_a}{d-1}|\lambda\rangle \Theta(1/d - \langle\lambda|P_a|\lambda\rangle) + \frac{\alpha_a}{d}(1 - \sum_{b=1}^{d^2} \alpha_b \langle\lambda|\frac{\mathbb{1} - P_b}{d-1}|\lambda\rangle \Theta(1/d - \langle\lambda|P_b|\lambda\rangle). \tag{E3}$$

Recall from the maintext that $E_a = \alpha_a P_a$, where $P_a$ are rank-1 projections. One can recognise that the first term in this response function is, upto a prefactor, given by the response functions for dichotomic measurements $\Theta(1/d - \langle\lambda|P_a|\lambda\rangle)$. The second term is constructed such that the response function is automatically normalised, $\sum_{a=1}^n G_a(\lambda) = 1$. It is easy to show that the function is positive, thus is a valid response function.

We need to compute the operator on the right hand side of equation (E2). To do this, we note

$$\int \mathrm{d}\omega(\lambda) G_a(\lambda) |\lambda\rangle\langle\lambda| = \alpha_a X_a + \frac{\alpha_a}{d}\left(\frac{\mathbb{1}}{d} - \sum_{b=1}^{d^2} \alpha_b X_b\right), \tag{E4}$$

where

$$X_a = \int \mathrm{d}\omega(\lambda) \frac{1}{d-1} \langle\lambda| Q_a |\lambda\rangle \, \Theta[\langle\lambda| Q_a |\lambda\rangle - (1-1/d)] |\lambda\rangle\langle\lambda|, \tag{E5}$$

where $Q_a = \mathbb{1}_B - P_a$. We again can show that $X_a$ is in the span of $\{\mathbb{1}, Q_a\}$, which can be characterised by

$$\mathrm{Tr}(X_a) = \frac{1}{d-1} a_1(d-1, 1-1/d), \tag{E6}$$

$$\mathrm{Tr}(X_a Q_a) = \frac{1}{d-1} a_2(d-1, 1-1/d). \tag{E7}$$

The critical value of $\eta_c$ where this construction of local hidden state model works is then

$$\eta_c = \frac{d^2}{d-1} a_2(d-1, 1-1/d) - d a_1(d-1, 1-1/d). \tag{E8}$$

With the explicit expressions of $a_2(d-1, 1-1/d)$ and $a_1(d-1, 1-1/d)$ one obtains equation (E1).

---


\end{document}